# Point singularity array with metasurfaces

Soon Wei Daniel Lim[1*†], Joon-Suh Park[1,2†], Dmitry Kazakov[1], Christina M. Spägele[1], Ahmed H. Dorrah[1], Maryna L. Meretska[1], Federico Capasso[1]

[1]Harvard John A. Paulson School of Engineering and Applied Sciences, 9 Oxford Street, Cambridge, MA 02138, USA

[2]Nanophotonics Research Center, Korea Institute of Science and Technology, Seoul 02792, Republic of Korea

*Email: lim982@g.harvard.edu

†Equal contribution

**Abstract**

Phase singularities are loci of darkness surrounded by monochromatic light in a scalar field, with applications in optical trapping, super-resolution imaging, and structured light-matter interactions. Although 1D singular structures, such as optical vortices, are the most common due to their robust topological properties, uncommon 0D (point) and 2D (sheet) singular structures can be generated by wavefront-shaping devices such as metasurfaces. Here, using the design flexibility of metasurfaces, we deterministically position ten identical point singularities in a cylindrically symmetric field generated by a single illumination source. The phasefront is inverse-designed using phase gradient maximization with an automatically-differentiable propagator. This process produces tight longitudinal intensity confinement. The singularity array is experimentally realized with a 1 mm diameter TiO$_2$ metasurface. One possible application is blue-detuned neutral atom trap arrays, for which this light field would enforce 3D confinement and a potential depth around 0.22 mK per watt of incident trapping laser power. Metasurface-enabled point singularity engineering may significantly simplify and miniaturize the optical architecture required to produce super-resolution microscopes and dark traps.

**Main text**

**Introduction**

Optical singularities occur when some parameter of the electric field is undefined; for instance, phase singularities occur when the wavefront phase is undefined at field zeros, and polarization singularities occur when at least one parameter of the polarization ellipse is undefined[1]. For random monochromatic scalar fields in a 3D space, such as in speckle patterns, 1D linear singularities (lines or curves) are ubiquitous since they are topologically protected against field perturbations. On the other hand, 0D (point) and 2D (sheet) singularities are far less common as they are not topologically protected. They tend to fragment into stable 1D linear singularities upon field perturbation[2], such as stray light either originating from external sources or deviations from the desired geometrical parameters of optical devices. Nevertheless, 2D singularities (membranes of darkness in 3D space) have been engineered and experimentally realized using wavefront shaping devices like metasurfaces[3]. Such devices can be obtained by inverse design optimization so that the light field achieves a large spatial gradient of the phase normal to the surface comprising the singularity.

While it is straightforward to position bright spots of light using conventional computer generated holography techniques such as Gerchberg-Saxton phase retrieval[4,5], these methods perform poorly at structuring dark regions of subwavelength dimensions[3]. 0D point singularities require the scalar field to vanish at only one point. These "cold spots" have been identified in the near-field of nanoparticles[6] and individual spots may be controllably displaced by superposing plane waves[7]. Here, we seek a method for deterministically placing multiple 0D singularities that is not bound to periodic spacing and does not mandate the use of multiple beams.

We present a straightforward method of deterministically positioning point singularities in a cylindrically-symmetric field. This strategy produces singularities with tight confinement, *i.e.*, small characteristic spatial dimensions with a rapid increase of the field intensity (amplitude modulus squared) away from the singularity point. We begin by describing the physical intuition behind the phase gradient maximization technique based on the geometrical structure of the 0D singularity. We then experimentally realize a linear array of ten tightly confined point singularities in the axial direction with a metasurface comprising $TiO_2$ nanopillars on glass. As a potential application, we evaluate the suitability of the resultant singular fields for neutral atom trapping in the blue-detuned regime, in which atoms are trapped in positions of darkness. While the engineered singularity array is very sensitive to the tilt of incident illumination, it is robust to wavelength changes of the trapping laser and demonstrates 3D confinement with no escape channels. Metasurface-enabled traps have the potential to greatly simplify the optical architecture required to produce dark optical traps for atoms or larger particles.

**Geometry of 0D singularities**

Point, *i.e.*, 0D, phase singularities occur when a complex scalar field $E$ is zero at only one point. That is, the real and imaginary zero-isosurfaces of $E$, loci of points for which $\text{Re}(E)=0$ and $\text{Im}(E)=0$, respectively, touch tangentially at only one point (**Figure 1a**). The field phase is defined for every point around the singularity except for the point itself (**Figure 1b**) and the intensity decreases quadratically to zero towards the singular point (**Figure 1c**). Similar to 2D sheet-like singularities, 0D point singularities are uncommon and fragile. They are not topologically protected and hence occur rarely in nature[3,8]. Nevertheless, they can be engineered to closely approximate 0D singularity behavior to within measurement uncertainties.

As with singularities of other geometries, the 0D singularity is accompanied by a region of large phase gradient magnitude $|\nabla\phi|^2=(\partial_x\phi)^2+(\partial_y\phi)^2+(\partial_z\phi)^2$ (**Figure 1d**) which diverges to infinity at the position of the singularity. In optical fields, this phase accumulation rate can be much larger than the field wavenumber $k=2\pi/\lambda$, indicating superoscillatory behavior[9].

While phase singularities can be engineered by enforcing perfect destructive interference at a point, the *confinement* of the dark point is another critical parameter, especially in superresolution microscopy (*e.g.*, STED[10]) and optical trapping. In these applications, dark positions should ideally be fully surrounded by light (*i.e.*, 3D confinement) with sharp field gradients (*i.e.,* tightly confined/localized). These additional constraints on the field distribution in the vicinity of a dark point cannot be satisfied by simply minimizing the field intensity at the target position of the 0D singularity. Here, we show that *phase gradient maximization* can enforce singular behavior at a point while simultaneously achieving tight confinement around the singularity. To build intuition for this technique, we first consider a complex field $E$ along a line, and compare the fields that are produced by a simple intensity minimization at *z=0* and a phase gradient maximization at that same point (**Figure 2**). To avoid plotting unrealistically high phase gradients, we show fields that have a finite minimum intensity $\epsilon>0$. Such a system may not yield zero intensity due to fabrication imperfections or optimization constraints. Optimization constraints arise when one seeks to balance multiple competing desired behaviors, *e.g.*, in a multi-objective optimization for which one simultaneously optimizes the field structure at different locations. Close to an intensity minimum, the real and imaginary field components ($E_r$ and $E_i$, respectively) are approximately linear and with opposite slope directions (**Figure 2a**). Since engineering the singularity by minimizing the field intensity at *z=0* just enforces a small $\epsilon$ there, it is insensitive to the slopes of $E_r$ and $E_i$ across the singularity, which can be shallow and thus produce a slowly varying field intensity minimum with weak localization. At *z=0*, $E_r$ and $E_i$ change sign and thereby produce a $\pi$ phase shift across the field minimum. This shift in phase can be captured by the variation in phase gradient $\partial_z\phi$, which has a broad and short peak at *z=0* (**Figure 2c**). Engineering a singularity by simply minimizing the intensity at the desired field minimum position does not give one control over the confinement there.

On the other hand, maximizing the phase gradient at *z=0* simultaneously achieves singular behavior and improves confinement. Intuitively, noting that the phase gradient can be written in terms of field gradients $\nabla\phi=\mathrm{Im}(\nabla E/E)$, maximizing $\nabla\phi$ not only minimizes the value of *E* in the denominator but also maximizes the field gradients $\nabla E$ in the numerator. This means that the slopes of $E_r$ and $E_i$ are steeper across the singularity, producing a more rapidly varying field intensity minimum with narrower spatial confinement (**Figure 2b**). A higher peak phase gradient also yields a taller and narrower phase gradient peak across the field minimum so that the accumulated phase across the minimum remains π (**Figure 2d**).

In three dimensions, 0D singularities are characterized by large phase gradients in all directions. One has to simultaneously maximize the phase gradients at the same point to "squeeze" the singularity into a point, a task which poses convergence difficulties since changing one directional gradient at a point inevitably affects the other gradients in the other directions. This problem is circumvented when the field is constructed to be azimuthally (cylindrically) symmetric about the optical axis: *i.e.*, the electric field $\boldsymbol{E}(r,z)$ is only a function of the radial distance from the optical axis $r$ and longitudinal position along the optical axis $z$. One can produce 0D point singularities along the optical axis just by maximizing one directional gradient at each of the desired points: the $z$-directed phase gradient. This exploitation of a system symmetry improves numerical convergence to an optimal design.

As a proof-of-concept for 0D singularity engineering, we designed an array of ten 0D singularities spaced 3 μm apart (**Figure 3a**) to be generated by a phase-only metasurface measuring 1 mm in diameter, and illuminated by a narrowband laser centered at λ=760 nm. Although we demonstrate a uniform array of singularities here, the algorithm can be applied to aperiodic singularity patterns as well, and we show one such design in **Supplementary Figure 1**. Such a light field, structured *longitudinally* along the optical axis, is challenging to generate using conventional holography methods that excel at designing only *transverse* field patterns. Full 3D holographic pattern generation with both transverse and longitudinal control remains an area of active research[11]. We partition the cylindrically symmetric metasurface plane into 1001 annular regions, each 500 nm thick. Each annular region is assigned a transmission phase delay so that the metasurface system can be parametrized by the 1001 phase delay values which serve as tunable optimization parameters. The phase profile from the metasurface is propagated into free space ($z>0$) using a vectorial propagator[12] built on an automatically differentiable platform (Tensorflow[13]), assuming that the incident field is initially linearly *x*-polarized for simplicity. The process is generalizable to optimizing both transverse polarization components over the surface and is not restricted to single scalar fields. This automatically differentiable propagator affords computationally efficient calculation of the exact numerical gradients of arbitrary objective functionals on the diffracted field.

There are two steps in the optimization process. In the first stage of optimization, we maximize the longitudinal phase gradient of the *x*-polarized $E_x$ field at ten equally-spaced target singularity positions from z=500 µm to z=527 µm along the optical axis. The radially-oriented phase gradient is identically zero due to azimuthal symmetry and continuity conditions for analytic fields: a nonzero radial phase gradient along the optical axis will produce a kink in the phase gradient across the optical axis. This first step produces a 0D singularity at each of the target positions. The intensity (*i.e.*, $|E_x|^2+|E_y|^2+|E_z|^2$) and $E_x$ phase profiles around each of the singular positions after this first step are plotted in **Supplementary Figures 2** and **3**, respectively. In several positions, the real and imaginary zero-isolines come close but do not touch, indicating that these situations are close approximations of 0D singularities and not mathematical 0D singularities. In the second stage of optimization, we use the optimized first stage result to equalize the phase gradient and second spatial derivative of $|E_x|^2$ (as a proxy for the intensity) over all the singularity positions and thus obtain nearly identical singularities across the array. The field intensity and phase profiles around each dark position are plotted in **Supplementary Figures 4-5**, respectively. The phase gradient profile of the $E_x$ field along the optical axis is plotted in **Figure 3b** and shows identical large superoscillatory values of $100k_0$ at the singularity positions, as designed. High spatial resolution plots of the phase gradients around each singularity position are shown in **Supplementary Figure 6**, which also show that the full-width-at-half-maximum of the phase gradient magnitude is 2.3 nm for each singularity. The tight feature localization of optical singularities has been exploited for precision displacement sensing[14]. The inverse-designed phase profile along the metasurface is unwrapped and plotted in **Figure 3c** to show the long-range structure. Full details of the optimization process are in **Supplementary Information section 1**.

We fabricated a metasurface comprising 700 nm tall cylindrical $TiO_2$ pillars in a 1 mm diameter metasurface on a fused silica substrate to enforce the required phase profile to generate ten 0D singularities. The fabrication process is similar to previously published work[15] and involves electron beam lithography of the required nanopillar profile into electron beam resist, followed by atomic layer deposition of amorphous $TiO_2$ into the developed resist voids. Excess $TiO_2$ is etched back using reactive ion etching to leave free-standing nanopillars. An opaque aluminum aperture is positioned around the metasurface to reduce stray light. Details of the nanofabrication process are in **Supplementary Information section 2** and the nanopillar library optical performance is plotted in **Supplementary Figure 7**. At each metasurface position, we pick the nanopillar from the library that has the closest transmitted phase to the required phase at that radial position. The non-uniform transmission amplitude of the meta-atom library introduces slight field deviations from the design field distribution, and we plot the predicted field intensity and phase profiles incorporating these imperfections in **Supplementary Figures 8-9**, respectively. The field intensity structure is largely preserved but the phase profile is slightly distorted near the intended singularity positions. These deviations can be avoided by incorporating the nonuniform transmission intensity of the nanopillar library during optimization. Scanning electron microscope images of the fabricated metasurface are shown in **Figure 3d**. For characterization, the metasurface is illuminated with a narrowband distributed feedback diode laser ($\lambda$=760 nm, 2 MHz linewidth) coupled to a single mode fiber with collimated output, and the transmitted field through the metasurface is captured over 1201 longitudinal *z*-positions at steps of 50 nm, where *z*=0 mm corresponds to the patterned surface of the metasurface, using a high magnification (100x, NA=0.95) horizontal microscope system (**Figure 3e**). The transmitted intensity measurements are normalized to the incident power flux at the metasurface. Full experimental and data processing details are in **Supplementary Information section 3**.

**Results and Discussion**

The simulated cylindrically symmetric field intensity profile on the *xz* plane in the vicinity of the ten 0D singularities is plotted in **Figure 4a**. The Cartesian polarization components are plotted separately in **Supplementary Figure 10**. The experimental intensity profiles in the longitudinal *xz* and *yz* planes are displayed in **Figure 4b** and **c**, respectively, and demonstrate good agreement with the simulated profiles. The intensity profile colormaps are adjusted to show the singularity region clearly and some parts of the surface plots are saturated. The maximum intensity value is indicated adjacent to each plot. The longitudinal cuts were obtained by stacking the 1201 captures of the transverse field intensity. The captured transverse *xy* field intensity at and between the ten singular positions are shown in **Figure 4d**, with rings of light around the dark singular points and bright on-axis spots in between singular positions. These transverse intensity pictures are stacked in the longitudinal direction to produce the *xz* and *yz* cuts in **Figure 4b** and **c**, respectively.

We observe that the experimental intensity is about a factor of four times smaller than the numerically predicted intensity. This is due to our intensity normalization choice and diffractive losses from the breaking of the ideal periodic boundary condition that underlies our metasurface library. We underestimate the field intensity by measuring the transmitted field power profile after it passes through the microscope objective and tube lens, thereby incorporating the reflective losses from multiple interfaces. We also overestimate the incident power by neglecting power loss due to Fresnel reflections off the fused silica-air interface.

Due to their high intensity gradients, phase singularities are effective as optical traps. Dielectric particles with a refractive index lower than the surrounding medium, reflective particles, and absorptive particles can all be trapped in the dark minimum of a beam, such as that on the axis of a donut beam carrying orbital angular momentum[16,17]. For neutral atoms, depending on the sign of the detuning $\Delta=\omega-\omega_0$ between the optical trap field frequency $\omega$ and a strong atomic resonance frequency $\omega_0$, such atoms are attracted to either intensity maxima ("red" $\Delta<0$ detuning) or minima ("blue" $\Delta>0$ detuning)[18]. Most optical dipole traps for neutral atoms are red traps which trap neutral atoms in arrays of tightly focused spots of light. Blue "bottle" traps with 3D spatial confinement, which trap the atoms in a dark spot surrounded by light, are more difficult to realize but provide several key advantages over red traps. Atoms trapped in blue traps experience substantially lower scattering rates[18] and thereby longer coherence times[19]. Importantly, the trap laser can remain on during laser excitation with other coherent sources[20].

Techniques using a single structured beam have been able to produce single blue traps[19,21–23], more exotic bottle traps based on acoustic[24,25] or ponderomotive[26] forces, and arrays of blue traps in the transverse plane[20,27]. The state-of-the-art blue trap array in active use is arguably the quantum gas microscope[28], which holographically projects a two-dimensional optical lattice into a vacuum cell, thereby achieving thousands of trap sites with individual optical access.

There is growing interest in using metasurfaces for the generation of atom traps[29–31], where the multifunctional, compact metasurface can replace multiple conventional optics and may even be located within the vacuum chamber. Recently, Hsu et al performed single atom trapping with a red detuned trap generated by a metalens inside the vacuum chamber[31].

The geometrical parameters of the 0D singularity array shown here are compatible with that of cold $^{87}$Rb Rydberg atom arrays[32] ($D_2$ line at 780.241 nm) and may conceivably be deployed in the orthogonal geometry portrayed in **Supplementary Figure 11b**, where a single metalens and single-sided illumination can generate the multiple blue traps for optical interrogation in the transverse direction. This is in contrast to the in-line architecture of optical traps in which the trapping and optical interrogation is performed through the same high numerical aperture objective. The trapping depth (in mK temperature units) per incident laser power is predicted to be 1.9 mK·W$^{-1}$ for the numerical simulation and 0.2 mK·W$^{-1}$ for the experimental intensity profile (**Supplementary Information section 4**). Both intensity profiles do not have any escape channels. **Supplementary Information section 5** evaluates the sensitivity of the structured optical field to changes in incident illumination tilt and incident wavelength on the metasurface. Although the light field is tolerant to changes in the incident wavelength on the order of 10 nm, the effective angular bandwidth is around 2 mrad (0.11°). This is similar to the field of view of 0.2° obtained in the previously reported metasurface red optical tweezer with NA=0.55[31]. This limited angular bandwidth may be overcome with metasurface angular dispersion engineering[33] to obtain better angular performance by trading off unneeded chromatic bandwith[34].

Passive metasurfaces excel in applications which afford very little volumetric and mass footprint while demanding high performance under a narrow set of constraints. The latter is due to the inherent trade-off between chromatic control, angular dispersion, and efficiency[34,35]. Given the space limitations in ultra-high-vacuum chambers and well-defined operational wavelengths for controlling and interrogating trapped particles in atomic physics, metasurfaces may be ideal for compact, few-component atom trap architectures. The 0D singularities generated by such metasurfaces are suitable for deployment as blue-detuned trap arrays and can also be accentuated in future work with dispersion engineering[36] to perform additional functions under illumination with different laser wavelengths or capture fluorescent emissions from the trapped atoms. Beyond optical traps, engineered 0D singularities may also be used in MINFLUX superresolution microscopy[37] to capture information simultaneously from multiple points.

## Acknowledgements

S.W.D.L. is supported by A*STAR Singapore through the National Science Scholarship Scheme. J.S.P. is on leave from the Korea Institute of Science and Technology. M.L.M. is supported by NWO Rubicon Grant 019.173EN.010, by the Dutch Funding Agency NWO. This material is based upon work supported by the Air Force Office of Scientific Research under award number FA9550-22-1-0243. This work was performed in part at the Harvard University Center for Nanoscale Systems (CNS); a member of the National Nanotechnology Coordinated Infrastructure Network (NNCI), which is supported by the National Science Foundation under NSF award no. ECCS-2025158. The computations in this paper were run on the FASRC Cannon cluster supported by the FAS Division of Science Research Computing Group at Harvard University. The authors thank Mikhail Lukin (Harvard University), Brandon Grinkemeyer (Harvard University), and Sooshin Kim (Harvard University) for helpful discussions.

**Author contributions**

S.W.D.L. conceived the algorithm and design with input from A.H.D. J.S.P. fabricated the samples. S.W.D.L., J.S.P., and D.K. designed the experiment and characterized samples. C.M.S. derived the connections to trap confinement and perturbation sensitivity. M.L.M. clarified applications of the system. S.W.D.L. wrote the manuscript with contributions from all authors. F.C. supervised the research.

**Competing interests**

S.W.D.L., J.S.P., A.H.D, M.L.M, and F.C. are the inventors on a relevant provisional patent application (application number: 17/579,460) owned by Harvard University. The authors declare no other competing interests.

**Data availability statement**

The data that support the findings of this study are available from the corresponding author upon reasonable request.

**Code availability statement**

The code that supports the findings of this study is available from the corresponding author upon reasonable request.

**Figures and tables**

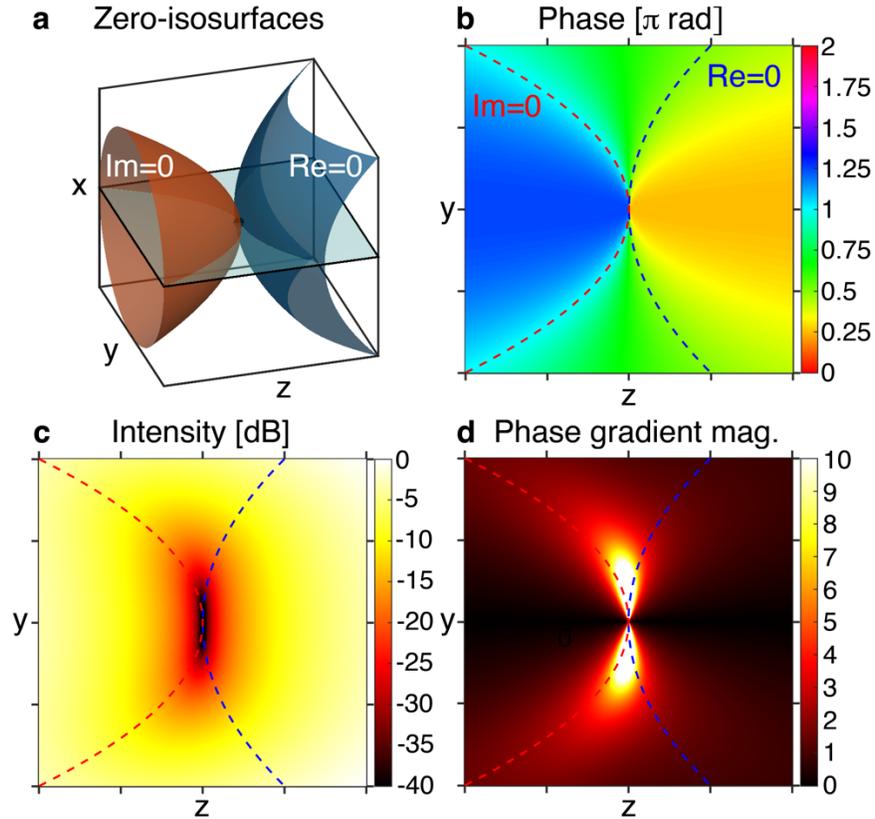

**Figure 1**. 0D singularity geometry. (**a**) 0D singularities in 3D space are isolated points of vanishing intensity in a scalar field *E*, occurring when the real (blue) and imaginary (red) zero-isosurfaces of *E* intersect tangentially. (**b**) *yz* cross-sectional phase and (**c**) intensity profiles of the 0D singularity in **a**. The dotted blue and red lines represent the real and imaginary zero-isolines of *E* on the plane, respectively. (**d**) Magnitude of the phase gradient $|\nabla\phi|$ in the *yz* plane, which is dominated by the minus *z*-directed phase gradient. The phase gradient diverges to infinity at the singularity position.

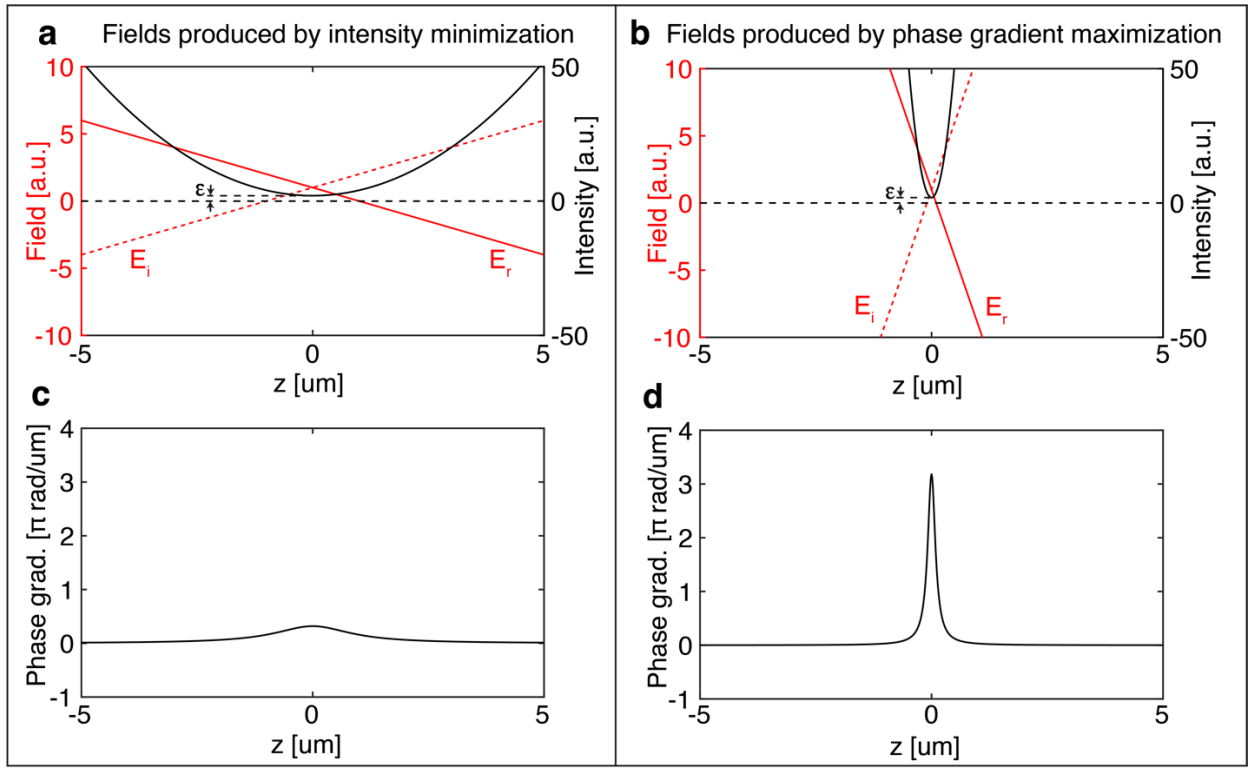

**Figure 2.** Comparison between two methods of producing 0D singularities: intensity minimization and phase gradient maximization. Only field behavior along the optic axis (*z* axis) is shown for simplicity. (**a**) Real ($E_r$) and imaginary ($E_i$) parts of scalar field $E$ in the vicinity of a low intensity position with minimum intensity $\epsilon$. Intensity minimization at *z=0* does not take the spatial distribution of fields around the low intensity point into account, producing fields with slowly varying $E_r$ and $E_i$ through the minimum, thereby producing a broad intensity minimum. (**b**) On the contrary, since phase gradient maximization *z=0* simultaneously minimizes the intensity there and maximizes the field slopes $\frac{dE_r}{dz}, \frac{dE_i}{dz}$ passing through that point, the resultant intensity minimum is narrow. (**c**) The phase gradient peak through *z=0* for the field in (**a**) produced by intensity minimization there is typically much lower than that of phase gradient maximization, as depicted in (**d**), which plots the phase gradient for the field in (**b**).

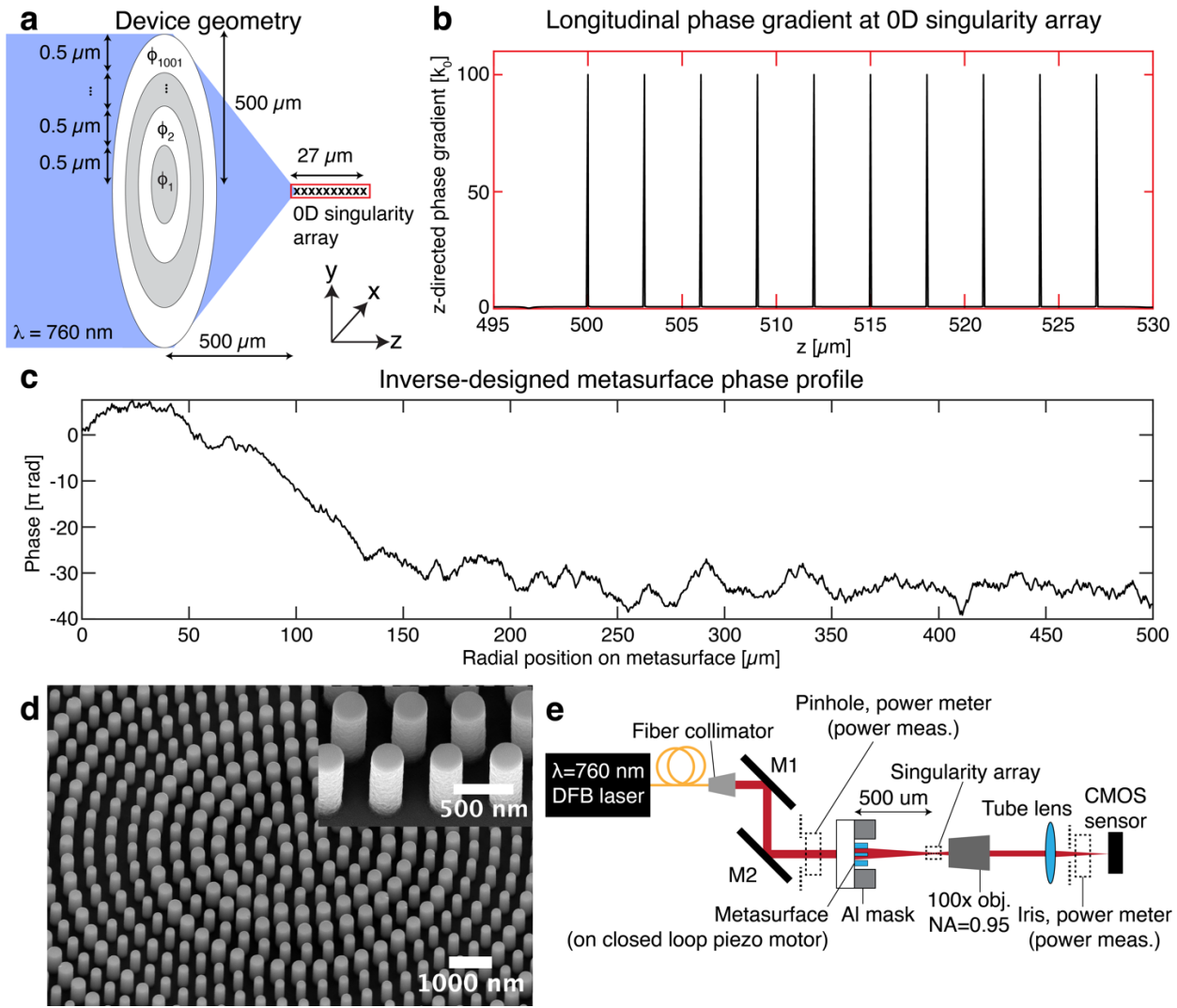

**Figure 3.** Design and experimental realization of 0D singularity array. (**a**) Geometry of the phase-only metasurface to generate the singularity array upon illumination by λ=760 nm light. The Cartesian directions are also indicated. (**b**) Longitudinal (*z*) phase gradient along the optic axis at the 0D singularity array, demonstrating large (compared to the free-space wavenumber $k_0$) and uniform phase gradients at the singularity locations. (**c**) Inverse-designed metasurface phase profile as a function of metasurface radial position that achieves the 0D singularity array. The phases have been unwrapped to show the long-range variation. (**d**) Scanning electron microscope image of the $TiO_2$ nanopillars on $SiO_2$ at the center of the fabricated metasurface that achieves the phase profile in (**c**). Inset: close-up of the nanopillars demonstrate vertical sidewalls. (**e**) Experimental setup to generate and characterize the 0D singularity array. Dotted lines indicate the positions of the pinholes and power meter used in characterizing the absolute transmission intensity.

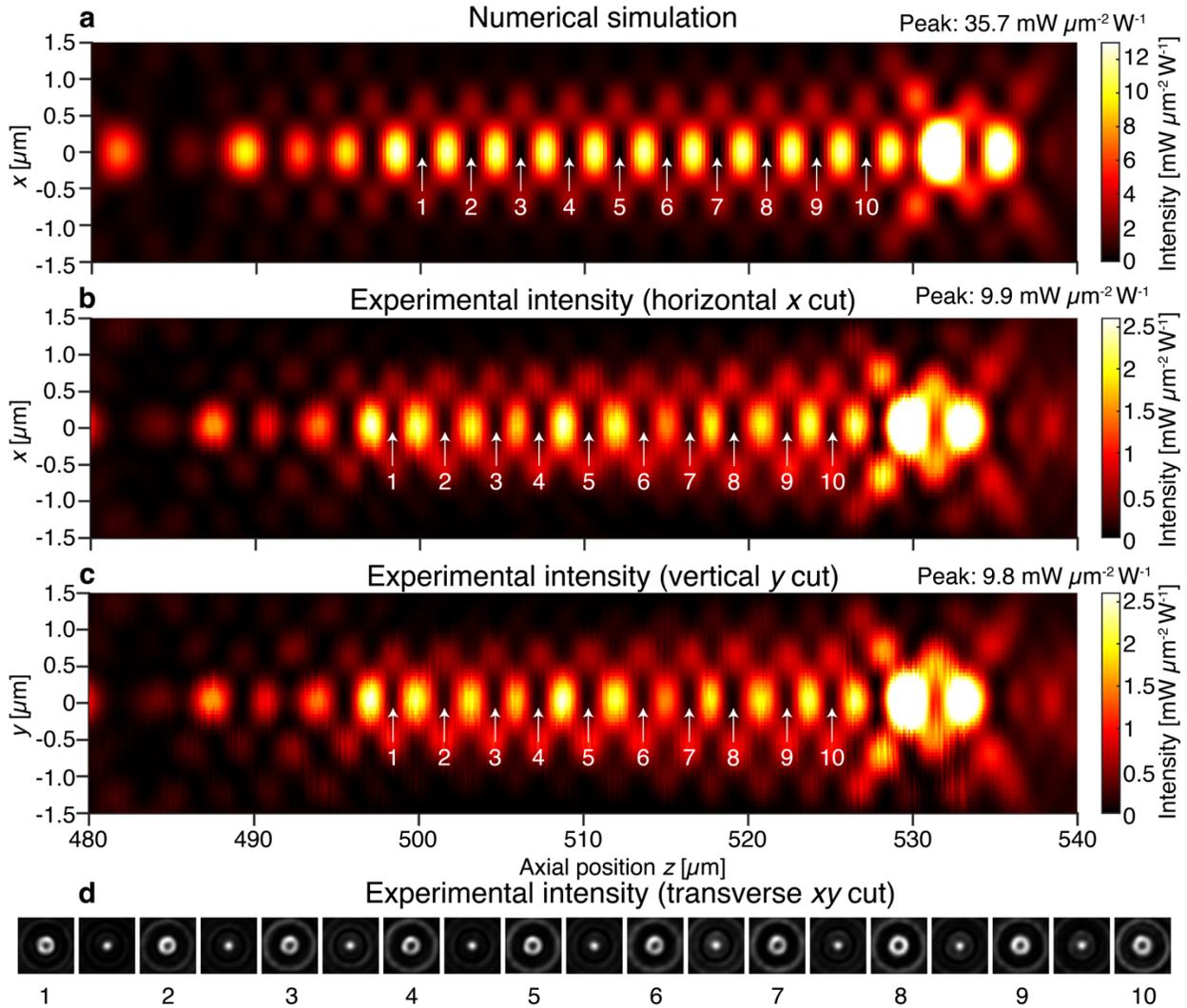

**Figure 4.** Longitudinal intensity cuts for singularity array with ten on-axis 0D singularities. The metasurface that produces this light field is located at *z=0*. The color scales are adjusted to show the singular region with higher contrast; peak intensity values for each of the colormaps are indicates in the top right-hand corner. White arrows indicate the locations of the ten 0D singularities. (**a**) Numerically simulated *xz* cut for the ideal metasurface. The *yz* cut is identical due to the rotational symmetry of the light field about the optic axis. (**b**) Experimental *xz* cut and (**c**) experimental *yz* cuts for the fabricated metasurface light field, demonstrating good agreement to the simulated light field. (**d**) Experimental transverse (*xy*) intensity profiles at each of the singularity positions (bright hollow annulus surrounding the 0D singularity) and in-between the singularity positions (focused spot). The longitudinal cuts in (**b-c**) are obtained by stacking 1201 such transverse images.

# Supplementary Information: Point singularity array with metasurfaces

Soon Wei Daniel Lim[1*†], Joon-Suh Park[1,2†], Dmitry Kazakov[1], Christina M. Spägele[1], Ahmed H. Dorrah[1], Maryna L. Meretska[1], Federico Capasso[1]

[1]Harvard John A. Paulson School of Engineering and Applied Sciences, 9 Oxford Street, Cambridge, MA 02138, USA
[2]Nanophotonics Research Center, Korea Institute of Science and Technology, Seoul 02792, Republic of Korea

*Email: lim982@g.harvard.edu
†Equal contribution

## 1. Computational design of metalens

The cylindrically symmetric phase-controlled metasurface at $z$=0 mm is parametrized by a set of 1001 annular rings, each of 500 nm radial extent, to produce a total lens with a 500 μm radius. For each radial position, we assign a scalar $\phi$ for the propagation phase delay of light there. This treats the metasurface as polarization-independent and cylindrically-symmetric. We then illuminate the phase-controlled surface with a uniform plane wave of vacuum wavelength 760 nm. This wavelength was chosen to be far blue-detuned from the D$_2$ resonance of Rubidium-87 atoms at 780.241 nm, which allows for the optical dipole potential to be written as directly proportional to the field intensity. This direct proportionality arises when the quantum-mechanical optical potential[1] is expanded to first order in the inverse detuning. We propagate this wavefront into the domain $z$>0 using the vectorial diffraction integral[2] implemented on an automatically differentiable platform (Tensorflow[3]). Automatic differentiation allows us to obtain objective function gradients efficiently and with a computational complexity that scales well with the number of degrees of freedom, which is 1001 in this problem. For concreteness, we assume that the incident polarization to the metasurface is oriented along the transverse $x$ direction. The $E_y$ and $E_z$ components vanish on-axis by cylindrical symmetry (**Supplementary Figure 10**). For intensity calculations off-axis, we use all three Cartesian components.

The 0D singularity positions are located at $r = 0$ and $z_i = 500$ μm, 503 μm, ... , 527 μm and are uniformly spaced 3 μm apart. The uniform spacing is not necessary; the positions can have any spacing along the longitudinal axis (**Supplementary Figure 1**). However, the light intensity between adjacent singular positions will become small if the spacing is comparable to or smaller than the characteristic longitudinal extent of focused spots, which can be approximated by the depth of focus $\lambda/NA^2$, where $NA$ is the equivalent numerical aperture for a focusing lens with a focal spot at that longitudinal position. In this system, the equivalent depth of focus given the system NA is approximately 1.5 μm, which is smaller than the desired singularity spacing. The singularity spacings in this case are chosen to be close to that of prior [87]Rb Rydberg atom arrays[4].

We design the 0D singularity array using two stages of numerical optimization. These optimizations are performed with respect to the 1001 phase values over the cylindrically-symmetric metasurface. In the first stage of optimization, we maximize the longitudinal phase gradient at the target singularity positions. This step produces a 0D singularity at each of the target singularity positions. In the second stage of optimization, we use the optimized first stage result to equalize the phase gradient and second spatial derivative of the field intensity over all the singularity positions and thus obtain nearly identical singularities across the array.

For the first optimization stage, at each singularity position, we compute the z-directed phase gradient of the field $\partial \phi/\partial z$. The objective function $F_1$ to be minimized is the negative minimum of the squares of the z-directed phase derivatives for each singularity position:

$$F_1 = -\min\{(\partial \phi/\partial z_i)^2\}_{i=1,\ldots,n}$$

To improve convergence, we use a smooth approximation to the minimum function, which is analytic instead of being piecewise continuous:

$$\min(a_1, \ldots, a_n) = \frac{\log[\sum_i \exp(-s \cdot a_i)]}{-s}, s = \frac{100}{\frac{1}{n}\sum_i |a_i|} > 0$$

The sum inside the argument of the logarithm is dominated by the term corresponding to smallest value of $a_i$. $s$ is a scale factor chosen to bring the input array values onto the same approximate scale and avoid numerical loss of precision during the computation of the exponential.

In the second optimization stage, the objective function $F_2$ to be minimized is the maximum of the deviations of the phase gradient to a large target phase gradient, set here to be 100 times the nominal field wavenumber $k_0$, plus penalty terms for differences in the second spatial derivative of the on-axis intensity $I(z) = |E_x(r=0, z)|^2$:

$$F_2 = \max\left(\frac{\partial \phi}{\partial z_i} - 100 k_0\right)^2_{i=1,\ldots,n} + c_1 \frac{\sigma\{\partial_z^2 I(z_i)\}_{i=1,\ldots,n}}{\mu\{\partial_z^2 I(z_i)\}_{i=1,\ldots,n}} + c_2 \frac{\sigma\{\partial_r^2 I(z_i)\}_{i=1,\ldots,n}}{\mu\{\partial_z^2 I(z_i)\}_{i=1,\ldots,n}}$$

where $\sigma$ refers to the population standard deviation and $\mu$ is the population mean. $c_1$ and $c_2$ are weight parameters that are chosen so as to bring the three terms in $F_2$ onto similar scales. We use a smooth approximation to the maximum function to improve convergence, which is analogous to the smooth approximation to the minimum function described earlier.

$$\max(a_1, \ldots, a_n) = \frac{\log[\sum_i \exp(s \cdot a_i)]}{s}, s = \frac{100}{\frac{1}{n}\sum_i |a_i|} > 0$$

We find that $F_2$ convergence can be improved by ramping up the target phase gradient from around $5k_0$ to $100k_0$ in the objective function. That is, using the converged results from step 1, we target a phase gradient of $5k_0$ in $F_2$, find a local optimization minimum, then repeat the process for a higher target phase gradient until we reach the target of $100k_0$.

In order to realize this optimized radial phase profile in a metasurface that operates in transmission, we seek to place a meta-atom at each radial position (spaced in the circumferential direction by the meta-atom pitch of 500 nm) to enforce the required phase at that radial position. This nanostructure is chosen from a library of meta-elements comprising nanopillars made of 700 nm-tall amorphous TiO$_2$ mounted on a substrate of fused silica. These meta-atoms are shown schematically in **Supplementary Figure 7(a)** and the dependence of the phase and transmission efficiency as a function of the nanopillar diameter is plotted in **Supplementary Figure 7(b)**. Meta-elements close to resonances are removed from the library. Nanopillars of diameter between 80 nm and 480 nm provide 2π phase coverage and are thus used to construct the metasurface based on the required phase profile.

2. **Nanofabrication**

The metasurface comprises TiO$_2$ nanopillars on a glass substrate (0.5 mm-thick JGS2-fused silica) and is fabricated using electron beam lithography, atomic layer deposition, and reactive ion etching processes[5,6]. The nanopillar pattern is written into 700 nm thick ZEP520A electron-beam resist (Zeon Specialty Materials Inc.) using a high-speed 50 kV electron-beam lithography system (Elionix HS-50) followed by develop process in cold o-xylene solution. The patterned holes are then conformally filled with amorphous TiO$_2$ through low-temperature atomic layer deposition process (Cambridge NanoTech Savannah) until the holes are completely filled. The over-deposited TiO$_2$ is etched back using reactive ion etching with CHF$_3$/Ar/O$_2$ mixture (Oxford PlasmaPro 100 Cobra ICP Etching System) until the resist layer is exposed. The residual resist is removed by a downstream plasma asher (Matrix Plasma Asher, Matrix Systems Inc.), which leaves free-standing TiO$_2$ nanopillars or nanofins. A 1.1 mm diameter opaque aperture is formed around the 1 mm diameter metasurface by photolithography using S1818 photoresist (Kayaku Advanced Materials Inc.), electron beam evaporation of 150 nm thick aluminum (Sharon electron beam evaporator), followed by a lift-off process via overnight immersion in Remover PG solution (Kayaku Advanced Materials Inc.).

3. **Experimental characterization of point singularity array**

**Figure 3d** shows optical and scanning electron micrographs (Zeiss UltraSEM) of a singularity array metasurface processed under identical conditions to the metasurface used for optical characterization, respectively. The metasurface used for optical characterization was not imaged in the SEM because this requires the irreversible deposition of a conductive metallic layer. The experimental setup for characterizing the singularity array metasurface is depicted in **Figure 3e**. A 760 nm single frequency distributed feedback (DFB) laser (TOPTICA Eagleyard GmbH) is driven with a constant current source (Newport 505 Laser Diode Driver) and kept at a constant temperature (Newport 325 Thermoelectric Cooler Driver). The single mode fiber-coupled output is collimated with a reflective collimator (Thorlabs RC12APC-P01) and is incident on the fused silica face of the metasurface. The metasurface *z*-position is controlled using a closed-loop piezo-motor stage with nm resolution (Attocube ECSx3030). The transmitted light is captured using a horizontal microscope system comprising a high NA objective (Olympus 100x MPLAPON NA=0.95), tube lens (Thorlabs TTL-180A) and CMOS camera (Thorlabs DCC1545M). The intensity image is captured over a range of longitudinal *z*-positions at steps of 50 nm, where *z*=0 mm corresponds to the patterned surface of the metasurface. At each *z*-position, the system is allowed to stabilize for 10 seconds before multiple intensity images are captured at different exposure times ranging from 0.05 ms to 163 ms. These multiple exposure images are later weighted by their respective exposure times and stacked to remove saturated pixels and produce a composite image with a large intensity dynamic range.

To set an absolute power scale for the transmitted light field, we measure the incident and transmitted energy flux. The incident power is measured through a 1 mm pinhole (Thorlabs P1000K) using a silicon power sensor (Thorlabs S120B). The 1 mm diameter is chosen to match the diameter of the metasurface. The transmitted energy flux is measured indirectly by estimating the power flowing through each pixel of the CMOS sensor at the axial plane of maximum on-axis intensity, which occurs at around $z = 530$ μm. This is measured by placing an iris and power meter head between the tube lens and the CMOS sensor. The iris is used to reduce the diameter of the light beam incident on the CMOS sensor so that it fits entirely within the sensor area. By making the approximation that the intensity recorded by each pixel on the CMOS sensor (with the power meter removed) is proportional to the energy flux through that pixel, we are thus able to relate the intensity distribution recorded on the CMOS sensor to the total power flux recorded by the power meter head. The CMOS sensor image is captured at a high intensity dynamic range using a range of exposure times from 0.08 ms to 245 ms to improve the estimation precision. The power flux through the maximum intensity pixel on the transverse plane is used to set the absolute power scale for the $z$-stack measurements in **Figure 4**. Note that the experimental intensity measurements are an underestimation of the true intensity values since the transmitted intensities are measured after the microscope objective and the tube lens, which introduce reflective losses.

4. **Numerical characterization of the point singularity array as a blue trap array**

The light distribution in the vicinity of each singular point in the experimental array is characterized by fitting the volumetric light distribution to second-order polynomials. The axial location of each singular point is first determined by fitting the on-axis intensity in a 1D window (1 μm full width) around each singular point and estimating the minimum intensity axial position using the fitted coefficients. The fitted quadratic coefficient provides an estimate of the curvature of the on-axis intensity profile. The transverse intensity profile at each singular point axial profile is then fitted to a 2D quadratic polynomial using a window width of 1 μm in both transverse directions. The fitted quadratic coefficients yield the intensity profile curvature in the transverse directions.

The measured intensity profile $I(\mathbf{r})$ can be converted to optical potential values $U(\mathbf{r})$ in the context of neutral atom dipole traps by the relationship[1]:

$$U(\mathbf{r}) = \frac{\hbar\delta}{2}\log\left[1 + \frac{\frac{I(\mathbf{r})}{I_{sat}}}{1 + \left(\frac{2\delta}{\Gamma}\right)^2}\right]$$

$\delta = \omega - \omega_0$ is the detuning of the trap frequency to the dipole resonance frequency, $I_{sat}$ is the saturation intensity, and $\Gamma$ is the natural linewidth of the dipole transition. For the $D_2$ line of $^{87}Rb$, $\omega_0 = c\left(\frac{2\pi}{780.241 \text{ nm}}\right) = 2\pi \cdot 384.230 \text{ THz}, \Gamma = 2\pi \cdot 6.0666 \text{ MHz}, I_{sat} = 2.50399 \text{ mW/cm}^2$ [7]. In the limit where $I(r) \ll I_{sat}\left[1 + \left(\frac{2\delta}{\Gamma}\right)^2\right]$ and $\left(\frac{2\delta}{\Gamma}\right)^2 \gg 1$, the argument of the logarithm can be Taylor expanded to first order so that the optical potential is linear in $I(r)$:

$$U(r) = \frac{\hbar\delta}{2} \frac{\frac{I(r)}{I_{sat}}}{\left(\frac{2\delta}{\Gamma}\right)^2}$$

The curvature of the optical potential around the singularity is directly connected to the effective spring constant of an atom placed at the center of the singular position:

$$k_i = \frac{\partial^2 I(r)}{\partial i^2}, i = x, y, z$$

The trap frequency in each direction can then be calculated from the spring constants and the atom mass $m$.

$$\omega_i = \frac{1}{2\pi}\sqrt{\frac{k_i}{m}}$$

The trap depth for each point singularity is computed by taking the optical potential difference between the optical potential at the center of the trap and that of the first potential peak encountered when moving away from the center of the trap. Since the trap is not spherically symmetric, this potential difference will depend on the direction at which one moves away from the trap center. We introduce a polar coordinate system with the polar axis aligned with the $z$-direction of the optical axis. The azimuthal angle is the direction corresponding to cylindrical symmetry. For each polar and azimuthal angle, we draw a line emanating from the trap center and pointing in that direction. We then compute the optical potential along that line and identify the potential barrier as the height of the first peak encountered relative to the potential at the trap center. We quantify the trap depth as a function of the polar angle $\theta$ with respect to the optic axis by computing the trap depth for the $2\pi$ azimuthal angle range with the same polar angle, then taking the minimum (worst case) trap depth. These potential depths are plotted as a function of $\theta$ in **Supplementary Figure 10c-d**. The trap depth is nonvanishing for all polar angles, indicating that every trap obtained numerically and experimentally has 3D confinement. The overall trap depth for each trap can then be associated with the minimum trap depth over all polar angles. This overall trap depth is at least 1.87 mK/W for the simulated optical potential profile and is measured to be at least 0.22 mK/W for the experimental optical potential profile. The units of the trap depth are chosen to be in temperature units (through division with the Boltzmann constant $k_B$) and are normalized to the incident power on the metalens.

5. **Incident tilt and chromatic dependence of singularity array**

In **Supplementary Figure 12**, we examine the chromatic and incident beam tilt dependence of the singularity array metasurface on trapping behavior, partitioning the performance based on the trap index within the array from 1 to 10. These studies are performed numerically using the metasurface geometry and material optical parameters and assume a fixed 1W incident trapping power over the 1 mm metasurface. Under large chromatic shift or incident tilt, several trap positions lose 3D confinement and are not plotted in **Supplementary Figure 12**. Changing the incident trap wavelength changes the realized phase and amplitude profile of the metasurface in a manner similar to that of diffractive optics: the effective focal length of the metasurface decreases as the wavelength increases, causing the structured intensity pattern to translate along the optical axis with minor perturbation. The minimum potential depth is largely dependent on the detuning from the $D_2$ line and increases with decreasing detuning (**Supplementary Figure 12a**), although the variation in trap depths increases at smaller detuning as well. The trap center scattering rate, which depends on the field intensity at the trap center, remains low with small variations over trap positions for further detuning, but becomes larger and with high variation for reduced detuning (**Supplementary Figure 12c**). The axial and radial trapping frequency chromatic dependencies which are obtained by quadratic fitting of the trap potential in a 1 μm diameter window around each trap to obtain the potential curvature, are plotted in **Supplementary Figure 12e** and **g**, respectively. The inter-trap variation in trapping frequency is minimized at the design wavelength. The video of the field structure and trap positions as a function of increasing wavelength is included as **Supplementary Video 1**.

The tilt dependence of the trap array is of interest because of the potential use of metasurface traps in tweezer arrays. Tweezer arrays are currently able to produce multiple red-detuned trap positions by diffracting an incident trap laser into multiple outgoing tilt angles using acousto-optic deflectors[4,8,9]. Each diffraction order has a beam tilt that controls the transverse displacement of the focal spot from the optical axis when imaged into the vacuum cell using a high NA objective. One may consider "duplicating" the blue trap array by illuminating the metasurface with a number of trap lasers at different tilt angles, thereby producing one copy of the array at different transverse displacements. Our metasurface is not designed for off-axis illumination and thus shows a rapid fall-off in potential depth as a function of incident beam tilt angle (**Supplementary Figure 12b**), falling to half the potential depth at an angle of 2 mrad (0.11°). The trap center scattering rate increases nonlinearly with the beam tilt (**Supplementary Figure 12d**). The axial trapping frequency remains relatively stable with tilt (**Supplementary Figure 12f**), but the radial trapping frequency falls off more rapidly and is the reason behind several traps losing 3D confinement (**Supplementary Figure 12h**). The video of the field structure and trap positions as a function of increasing beam tilt is included as **Supplementary Video 2**.

**Supplementary Figures**

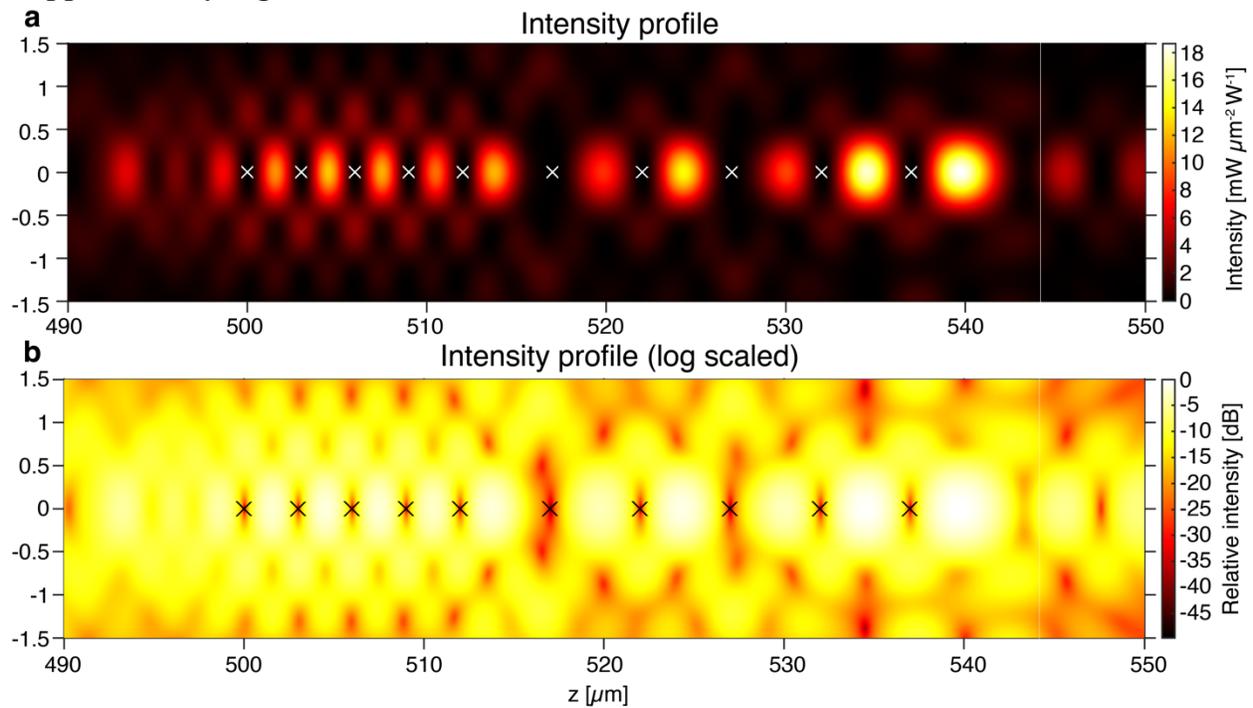

**Supplementary Figure 1.** Demonstration of a non-uniformly-spaced 0D singularity array. **a** *xz* intensity plot of a cylindrically symmetric 0D singularity array with five singularities spaced 3 µm apart (z = 500 µm to z = 512 µm) and five singularities spaced 5 µm apart (z = 517 µm to z = 537 µm). Crosses indicate the positions of the 0D singularities at which the phase gradient optimization was performed. **b** Logarithmically-scaled intensity plot of **a**.

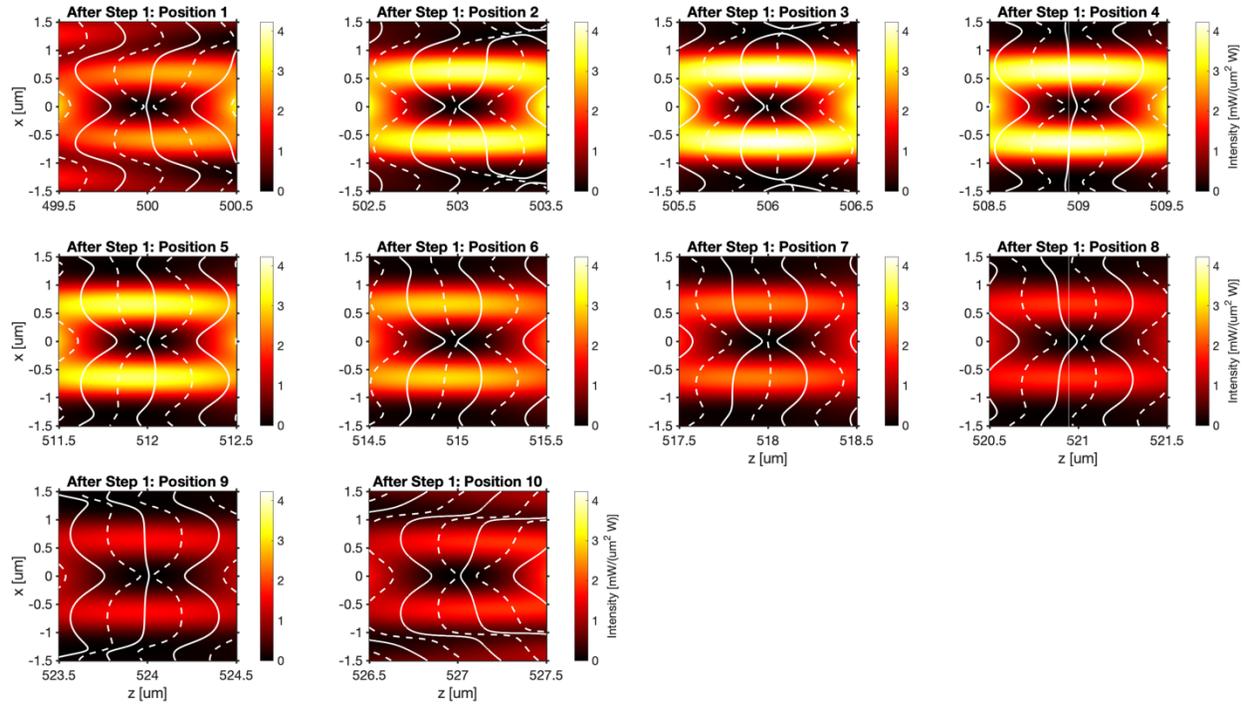

**Supplementary Figure 2**. Intensity $|E_x|^2+|E_y|^2+|E_z|^2$ profile and zero-isolines for the ten 0D singularity positions in the array, just after the first optimization step. Zero-isolines for the real part of the scalar field where Re($E_x$)=0 are plotted as solid black lines; zero-isolines for the imaginary part of the scalar field where Im($E_x$)=0 are plotted as dashed black lines. The amplitude profile at the metasurface plane is assumed to be uniform in this calculation.

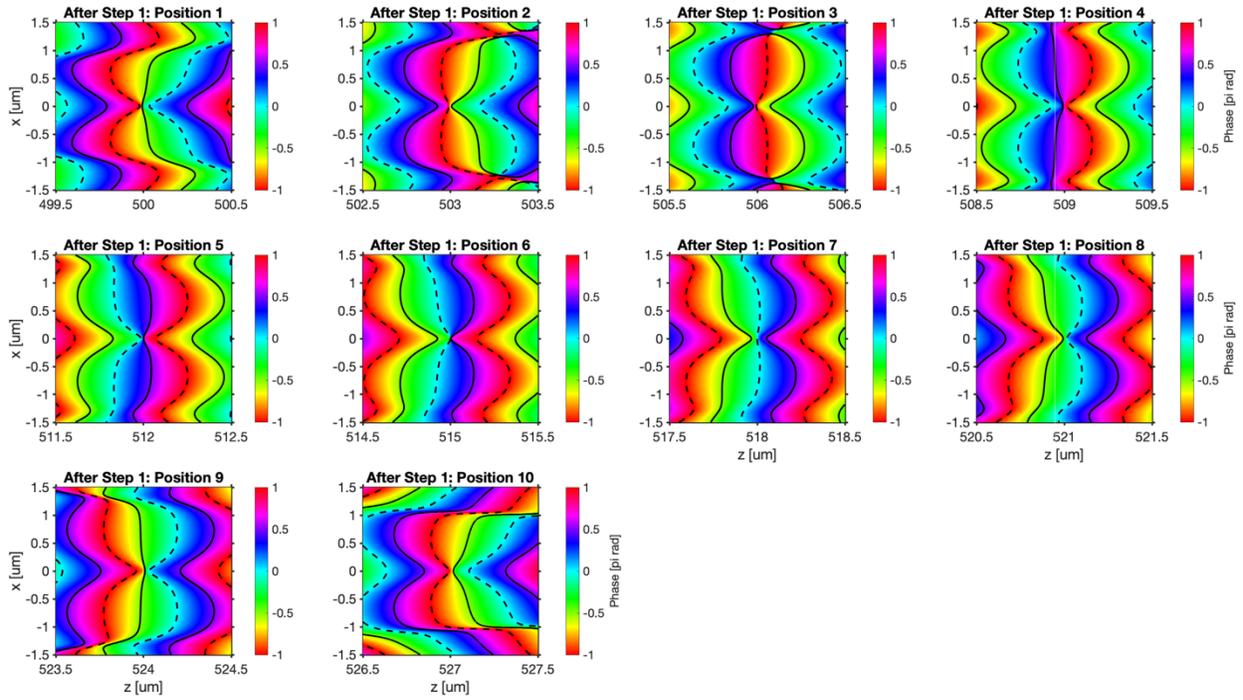

**Supplementary Figure 3**. $E_x$ phase profile and zero-isolines for the ten 0D singularity positions in the array, just after the first optimization step. Zero-isolines for the real part of the scalar field where $\text{Re}(E_x)=0$ are plotted as solid black lines; zero-isolines for the imaginary part of the scalar field where $\text{Im}(E_x)=0$ are plotted as dashed black lines. The amplitude profile at the metasurface plane is assumed to be uniform in this calculation.

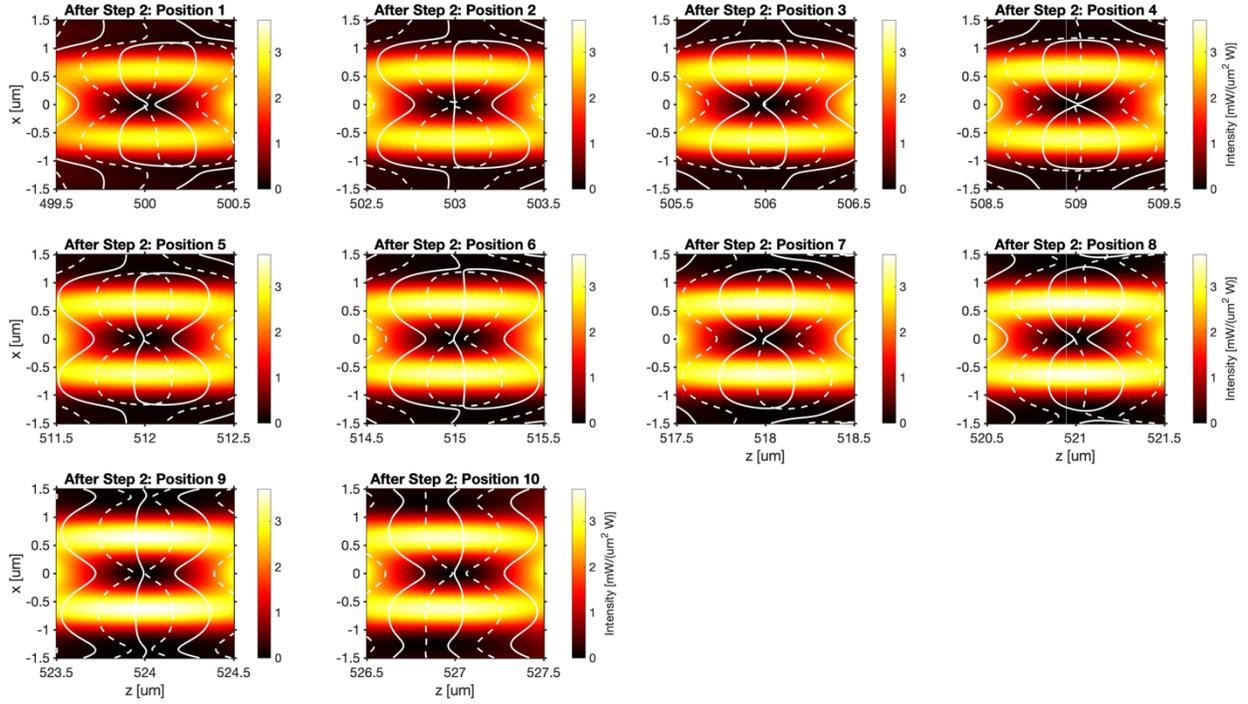

**Supplementary Figure 4**. Intensity $|E_x|^2+|E_y|^2+|E_z|^2$ profile and zero-isolines for the ten 0D singularity positions in the array, just after the second optimization step. Zero-isolines for the real part of the scalar field where $\text{Re}(E_x)=0$ are plotted as solid black lines; zero-isolines for the imaginary part of the scalar field where $\text{Im}(E_x)=0$ are plotted as dashed black lines. The amplitude profile at the metasurface plane is assumed to be uniform in this calculation.

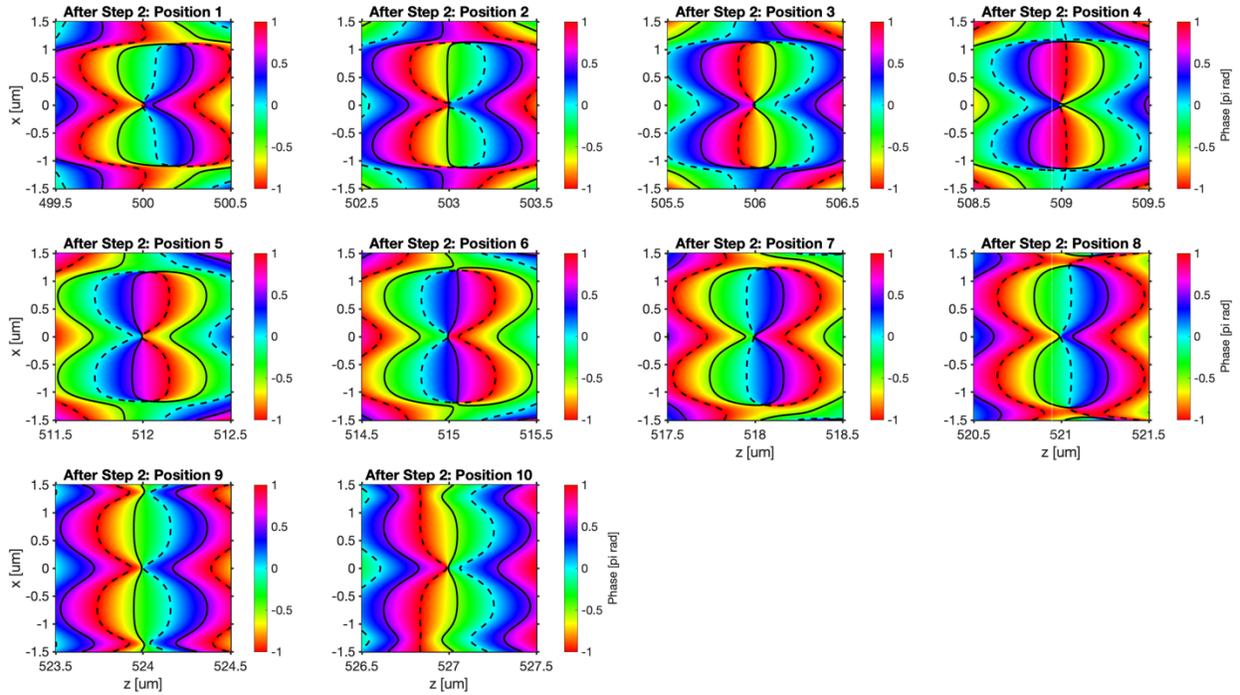

**Supplementary Figure 5**. $E_x$ phase profile and zero-isolines for the ten 0D singularity positions in the array, just after the second optimization step. Zero-isolines for the real part of the scalar field where Re($E_x$)=0 are plotted as solid black lines; zero-isolines for the imaginary part of the scalar field where Im($E_x$)=0 are plotted as dashed black lines. The amplitude profile at the metasurface plane is assumed to be uniform in this calculation.

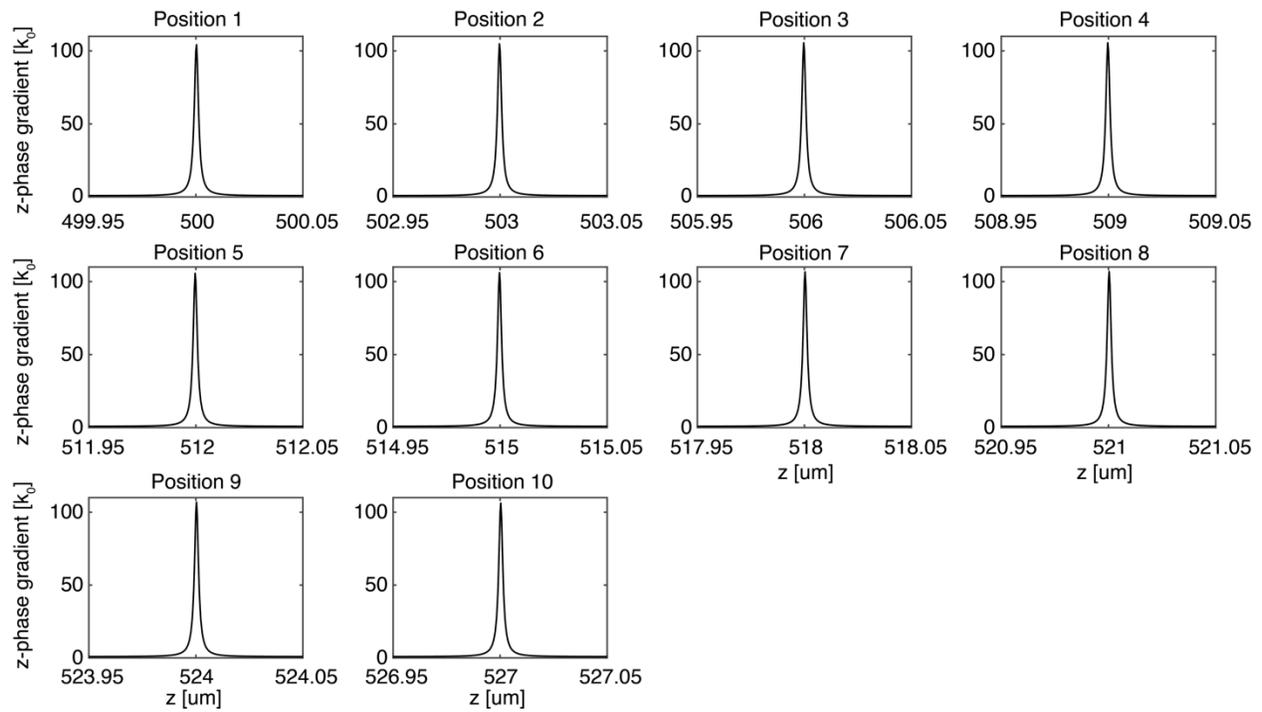

**Supplementary Figure 6**. Numerically-calculated $z$-directed phase gradient in the vicinity of each of the ten singularity locations. The full-width-at-half-maximum of the phase gradient is 2.3 nm for every location.

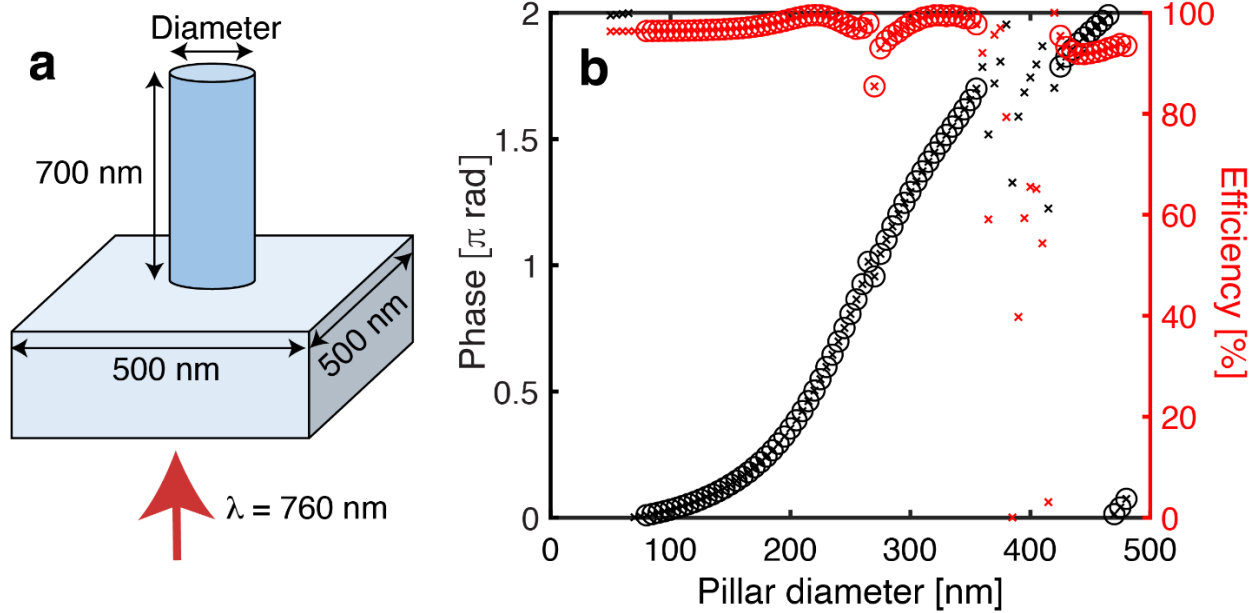

**Supplementary Figure 7.** **(a)** Cylindrical meta-atom geometry used in realizing the 0D singularity array. **(b)** Transmission phase and efficiency dependence on the nanopillar diameter for the cylindrical meta-atom. The circled data points are used in the meta-atom library. The diameter range used is 80 nm to 480 nm, which provides $2\pi$ phase coverage.

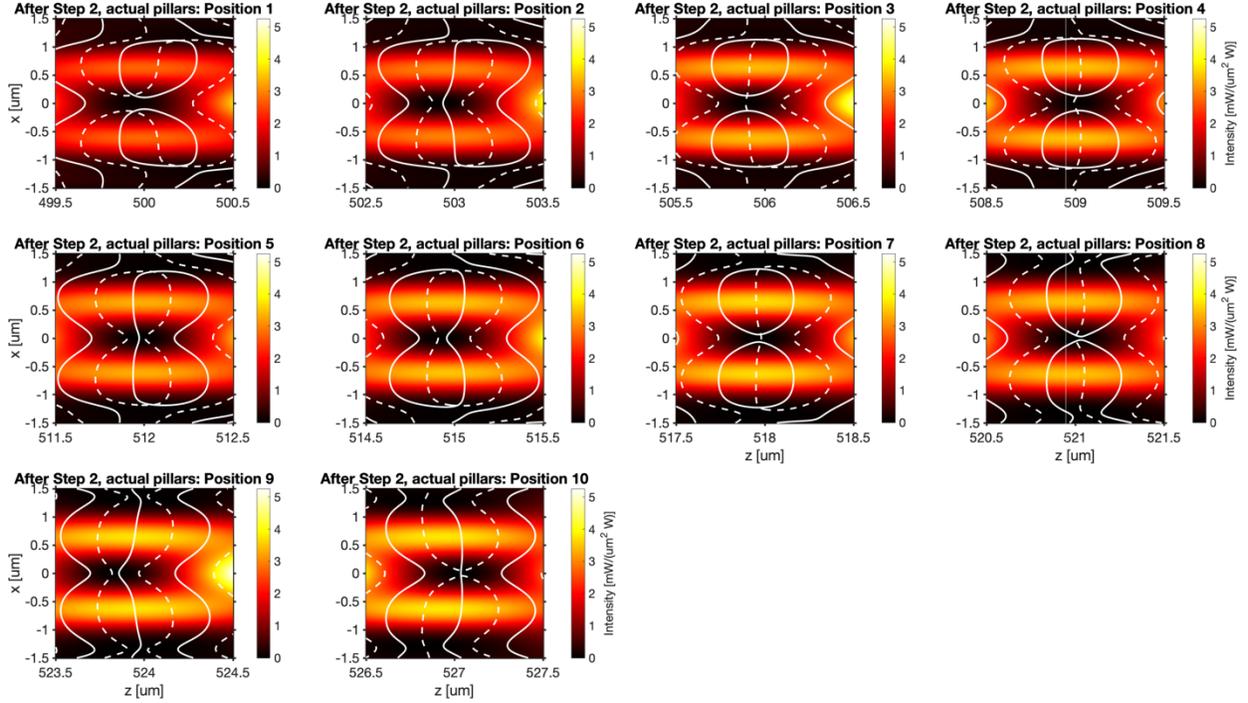

**Supplementary Figure 8**. Intensity $|E_x|^2+|E_y|^2+|E_z|^2$ profile and zero-isolines for the ten 0D singularity positions in the array, after the second optimization step, and incorporating the non-uniform transmission amplitudes of the TiO$_2$ nanopillar library. Zero-isolines for the real part of the scalar field where Re($E_x$)=0 are plotted as solid black lines; zero-isolines for the imaginary part of the scalar field where Im($E_x$)=0 are plotted as dashed black lines.

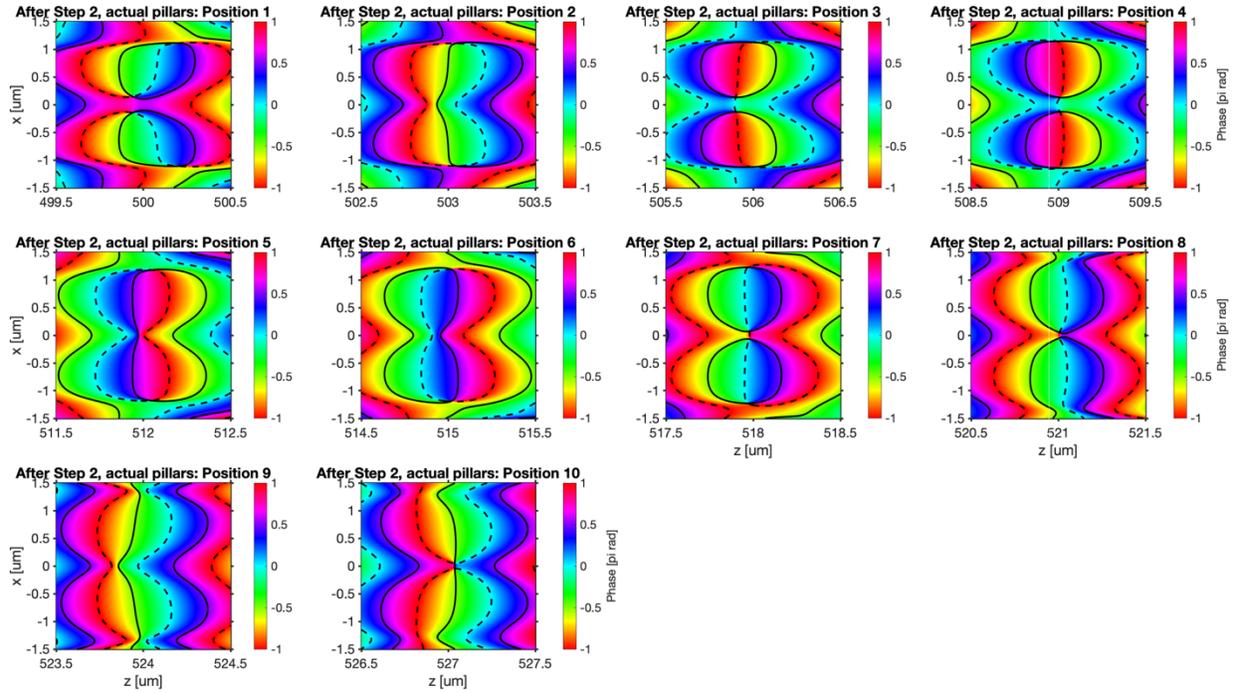

**Supplementary Figure 9**. $E_x$ phase profile and zero-isolines for the ten 0D singularity positions in the array, after the second optimization step, and incorporating the non-uniform transmission amplitudes of the $TiO_2$ nanopillar library. Zero-isolines for the real part of the scalar field where Re($E_x$)=0 are plotted as solid black lines; zero-isolines for the imaginary part of the scalar field where Im($E_x$)=0 are plotted as dashed black lines.

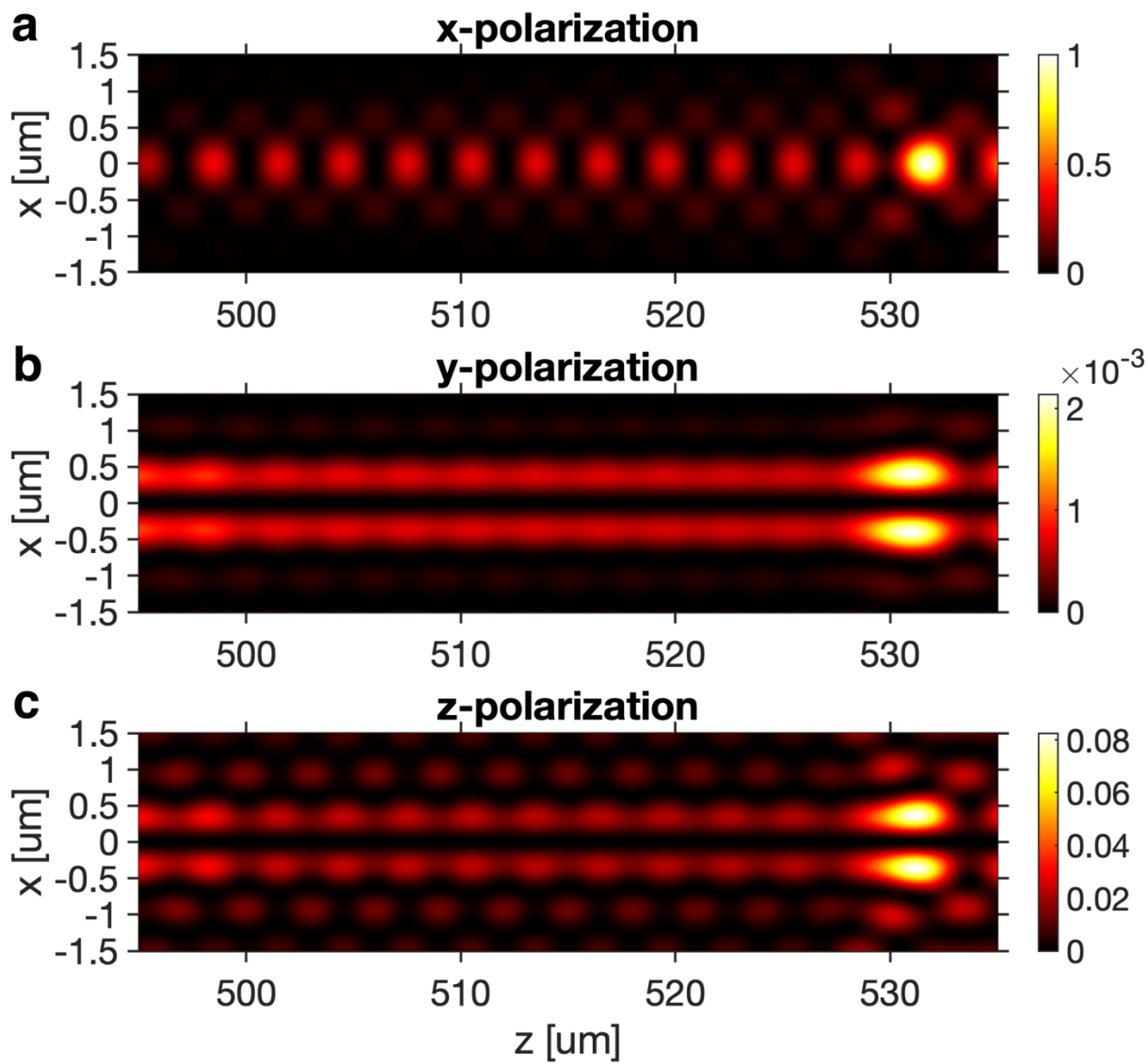

**Supplementary Figure 10 |** Simulated intensity contributions from each of the three Cartesian polarization components for the 0D singularity array. Each contribution is normalized to the maximum $E_x$ intensity. The total intensity is the sum of each of the contributions. The transverse polarizations are the (**a**) *x*-directed and (**b**) *y*-directed components. The longitudinal polarization is *z*-directed and is plotted in (**c**).

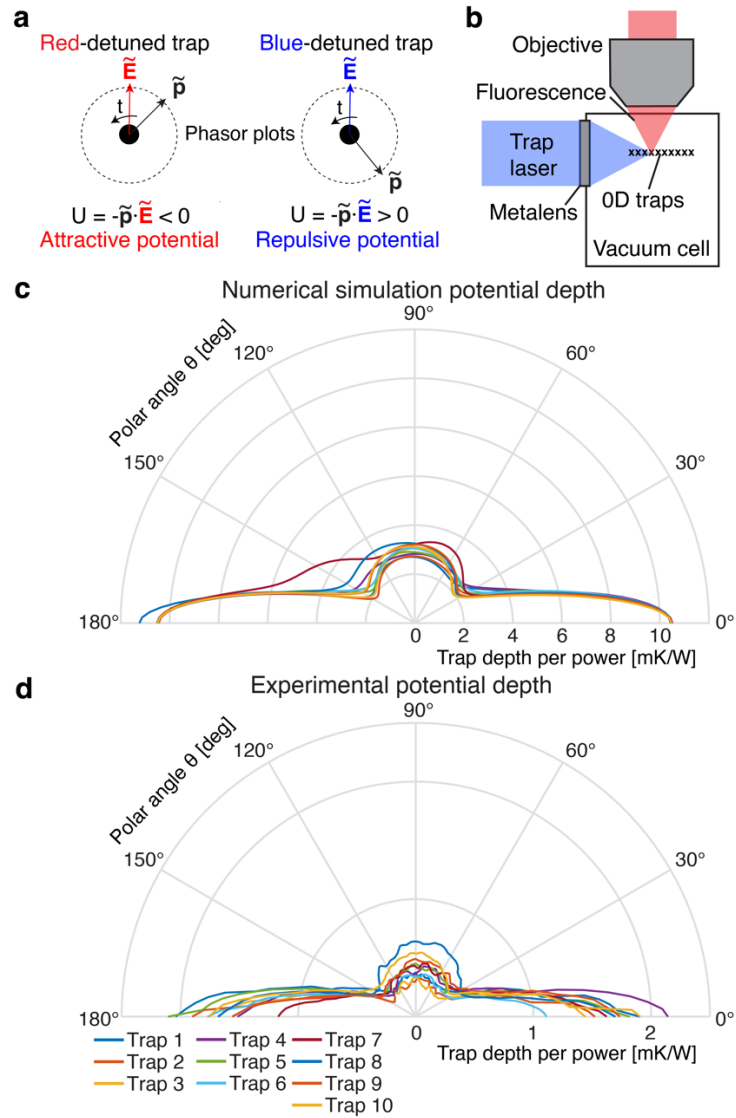

**Supplementary Figure 11.** Operation of 0D optical singularities as blue-detuned atom traps. (**a**) Time-domain phasor plots of the driving electric field **E** and driven electric dipole moment **p** for driving frequencies that are red and blue detuned from the dipole resonance. As time progresses, the phasors rotate counterclockwise. The phase angle magnitude between **E** and **p** for red-detuned frequencies is always less than $\pi/2$, leading to an attractive electric dipole potential. The phase angle magnitude is larger than $\pi/2$ for blue-detuned frequencies, leading to a repulsive potential. A dark spot surrounded by blue-detuned light serves as a blue trap for neutral atoms. (**b**) Possible vacuum cell configuration to trap and interrogate atoms that are trapped by the light field from a metasurface. (**c**) Numerically simulated potential depth for $^{87}$Rb atoms placed at the 0D singular locations of the simulated light field in **Figure 4(a)**. The polar angle $\theta$ is the angle from the optical axis and the potential depth is expressed in temperature units of millikelvin per watt of incident light on the metasurface. (**d**) Potential depth for $^{87}$Rb atoms placed at the 0D singular locations of the experimental light field in **Figure 4(b-c)**.

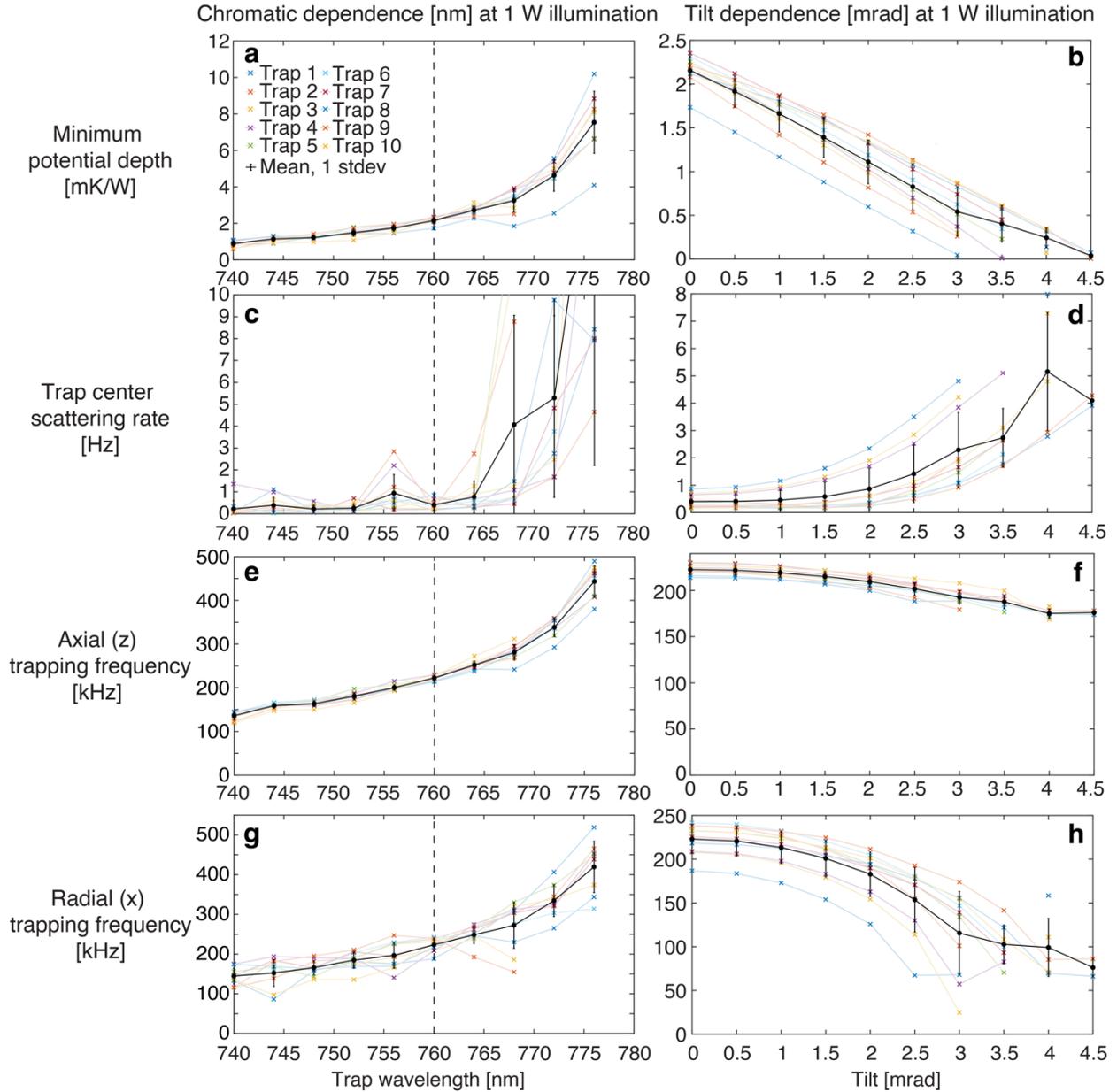

**Supplementary Figure 12**. Dependence of trapping parameters on wavelength and incident wavefront tilt for each of the ten 0D singularity traps generated by the metasurface used as blue traps for $^{87}$Rb. All calculations assume 1W of incident power at the metasurface. Error bars indicate one standard deviation of variation over the ten traps. (**a**) Escape potential depth, (**c**) trap center scattering rate, (**e**) axial and (**g**) radial trapping frequencies as a function of incident wavelength. (**b**) Escape potential depth, (**d**) trap center scattering rate, (**f**) axial and (**h**) radial trapping frequencies as a function of incident tilt. The tilt is defined as the angle from the surface normal for the incident wavefront at the air/glass interface on the back face of the metasurface.

**Supplementary Videos**

1. **Supplementary Video 1**: Variation of 0D singularity array field structure with changes in incident wavelength, linear scaled (top) and logarithmically scaled (bottom). Trap positions are indicated with crosses and vanish when 3D confinement is lost.
2. **Supplementary Video 2**: Variation of 0D singularity array field structure with changes in incident beam tilt on the metasurface, linear scaled (top) and logarithmically scaled (bottom). Trap positions are indicated with crosses and vanish when 3D confinement is lost.